\documentclass{article}
\usepackage{spconf,amsmath,graphicx}

\title{PROGRESSIVE HOLOGRAM GENERATION BASED ON OBJECT SALIENCY}
\name{Shima Rafiei
, Shahram Shirani}
\address{ \small Department of Electrical and Computer Engineering, McMaster University, Hamilton, Canada}
\begin{document}
%
\maketitle
\begin{abstract}
Computer-generated hologram (CGH) is promised to realize the next generation of 3D visual media with life-changing applications. However, one of the essential obstacles to this technology is the time-consuming hologram computation.
Thus, facilitating the computation of the generated hologram is of significant importance in this area. We propose a progressive hologram generation based on object saliency using discrete wavelet transform. In our method, the object is decomposed into 3 levels of resolution using wavelet transform.
Then, based on the saliency of the object, a progressive resolution hologram is generated. Our model generates a low-resolution hologram for non-salient areas of an object and a high-resolution hologram for salient areas, thus reducing CGH generation time. We applied our method to a number of objects and show that salient parts are reconstructed with high quality while it helps the process to speed up. Finally, we compare the SSIM of the reconstructed objects.


\end{abstract}
\section{Introduction}
\label{sec:intro}

Holographic images are able to fulfill all human visual systems' depth cues when displaying an object. Previously, obtaining holographic images was in need of an optical setup. However, to facilitate the process, a computer-generated hologram (CGH) digitally generates and reconstructs the hologram of an object using numerical calculation \cite{CGHIntro}. 
The object can be represented in a number of formats such as point cloud, light field, or images plus depth maps \cite{CGHIntro}. CGH Hologram generation can be summed up into two main approaches: point-source \cite{Frensel_basics} and wave-field method \cite{wavefield}. The point-source approach is based on Huygens–Fresnel principle \cite{Frensel_basics}. It considers each object point as a source of light whereas the wave-field approach benefits from light propagation approximation by calculating the propagation of light for an object layer at a single depth \cite{AngularSpec}. Point-source hologram generation has higher accuracy in generating exact fringe patterns on the hologram plane but is time-consuming while the wave-field approach is time efficient with the trade-off of losing the quality. \cite{hybrid} benefits from a hybrid method for hologram generation of 3D objects. It uses the point-source approach for object layers that consist of a low number of points, less than a threshold, and a wave-field approach for the rest of the layers to compensate for the computation time. Although their results show the benefit of using the point-source approach for some parts of the scene, choosing these points based on their frequency is a greedy approach. 
This paper also uses a look-up table (LUT) to load the pattern of the hologram for every point instead of calculating the Fresnel diffraction formula each time and then multiplying them into the intensity of the object point \cite{ganHOLO}. In hologram generation, several methods use LUT to reduce time computation \cite{LUT,videoLUT} and improved it via GPU implementations.  
\\
Considering the fact that hologram generation is computationally expensive in time, in this paper we show our hologram generation algorithm can enhance the trade-off between quality and time by utilizing knowledge about the object. Generally speaking, existing methods struggled to enhance the quality and time at the same time. They mostly do not apply their method to a complex object scene and also they treat all object points the same way. On the other hand, they do not detect the presence and location of an object before generating a hologram. For instance, recent approaches that utilize end-to-end deep learning methods mostly are capable of generating holograms for a specific and simple type of object \cite{Deep1,Deep2,Deep3}, i.e. they are not general-purpose. Our proposed method can generate holograms for any scene containing a variety of complex objects, thanks to PoolNet saliency detection \cite{PoolNet}. PoolNet is a deep convolution neural network that is a real-time salient object detector\cite{PoolNet}. This network incorporates aggregation modules to fuse the coarse-level semantic features with the fine levels, leading to a saliency map for the object with accurate border details. The saliency map for a whole scene helps us in the process of hologram generation. 
Additionally, we benefit from the signal expansion principle \cite{Gonzalez}, "any signal can be decomposed into a linear combination of basis". Using this principle, we consider our object, as a signal and decompose it into a summation of signals using Haar wavelet bases. Hence, we need to apply prepared hologram fringe patterns of different block sizes during hologram construction, to help the superposition principle works on the hologram side when the object size gets smaller.
Consequently, not only do we have a faster resultant hologram plane due to having smaller objects at each level, but also we develop a hologram plane into a higher quality only on blocks where the saliency maps emphasize the importance of an area. 

To sum up, our contributions are as follows:\\
(1) Proposing a progressive hologram generation using discrete wavelet transform (DWT).\\
(2) Speeding up hologram generation based on DWT and saliency map.\\
The remaining of this paper is organized as follows. In
section 2, we describe the proposed method along with qualitative results. In section 3, we express how we measure our method quantitatively and in section 4 we conclude the paper.

\section{Proposed method}
\label{sec:pagestyle}
CGH uses object wave equation, eq.\eqref{eqn:frensel} based on Huygens–Fresnel diffraction, to calculate hologram fringe patterns for object point sources \cite{objectwave,Antonin}. This formula shows the hologram $O(x,y)$ for $M$ object points with amplitude of $A_i$ at location $(x_i,y_i,z_i)$ with a wave number of $k$ and oblique distance $r_i$ from the hologram plane \cite{Frensel_basics}. 

\begin{equation}
\begin{aligned}
\centering
\label{eqn:frensel}
&O(x, y) = \sum_{i=1}^M{\frac{A_i}{r_i}}  \exp[j(k r_i)],
\\
&r_i = \sqrt{(x-x_i)^2 + (y-y_i)^2 + (z-z_i^2)}
\end{aligned}
\end{equation}

We know an object A can be decomposed into wavelet Haar bases of $\phi_j(x,y,z)$ and corresponding expansion coefficients of $\alpha_i$ \cite{Gonzalez}. In fact, as eq. \ref{eqn:wavelet_frenel} shows, at the first level of decomposition (L), we have four components of $L$, $H$, $V$, and $D$. In the same way, $L$ can be decomposed into the second level of DWT (LL) i.e. $LL$, $LH$, $LV$, and $LD$.

\begin{equation}
\begin{aligned}
\centering
\label{eqn:wavelet_frenel}
& \; \; \; \; \;\;\;\;\;\; \;\;\;\; \;\; \;\;\;\; A = \sum_{j}{\alpha_j{\phi_L}_j(x,y,z)}=
\\
& \overbrace{\sum_{j}{\alpha_j{\phi_{LL}}_j(x,y,z)}}^\text{L(x,y,z)}+ H(x,y,z)+ V(x,y,z)+ D(x,y,z) \\
\end{aligned}
\end{equation}



\begin{figure}[hbt!]
    \centering
    \includegraphics[scale = 0.35]{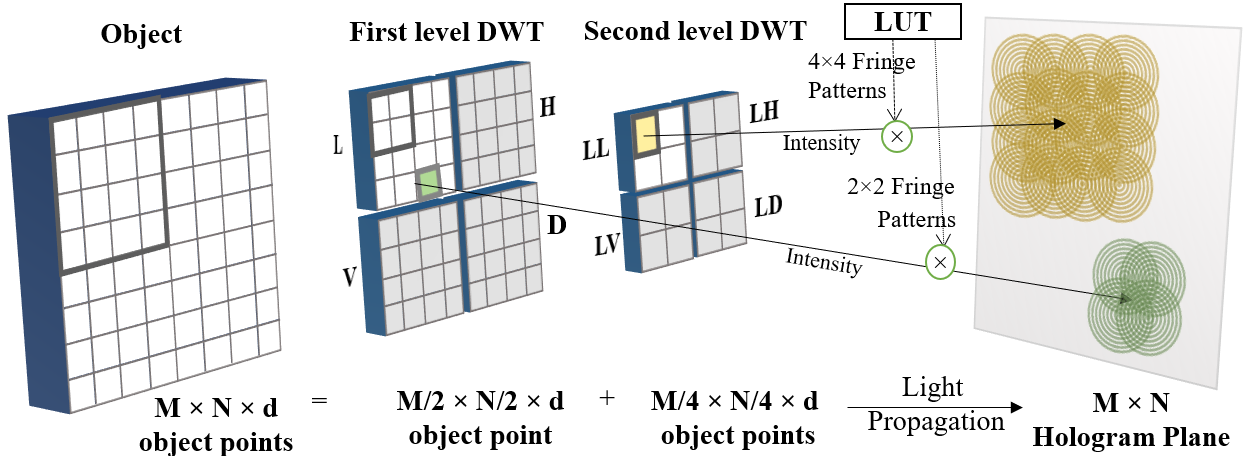}
    \caption{Superposition of fringe patterns on the hologram plane}
    \label{fig:superposition}
\end{figure}

If we rewrite eq. \ref{eqn:frensel} as object wave applied on the object A, to generate hologram $O(x,y)$, meaning that,
$${Object Wave(A) \rightarrow O(x,y)}$$

we can define $O(x,y)$ as,

\begin{equation}
\begin{aligned}
\label{eqn:arrow2}
&O(x,y) = \sum_{l}{\sum_{j}{ \: ObjectWave(\alpha_j \: {\phi_l}_j(x,y,z))}} \\
\end{aligned}
\end{equation}
which implies that the hologram of an object equals the sum of holograms of its Haar basis functions multiplied by the coefficient. Hence, we calculate hologram for the object's Haar basis instead of the object. Fig. \ref{fig:superposition} shows the superposition principle on the hologram plane. As shown in Fig. \ref{fig:superposition}, the object is decomposed into 2 levels of wavelet. In fact, for each level of DWT, we fetch the corresponding hologram block from LUT, multiply it by the intensity of that object point and then add it to the hologram plane.  
Here, yellow fringe patterns come from level 2, and green comes from level 1. For instance, each object point of $LL$  refers to the field of view of a $4\times4$ object point in the original image. Thus, in order to help the superposition theorem work correctly on the hologram plane, a $4\times4$ block of hologram has to spread the intensity of that object point onto 16 fringe patterns in the right position on the hologram plane. 
\\In this paper, we use three levels of wavelet and consider 2D objects. Therefore, we need hologram blocks of $8\times8$, $4\times4$, $2\times2$, and $1\times1$ precomputed in LUT using eq. \eqref{eqn:frensel}. The specification of our hologram computation for the object is shown in table \ref{table:specification}.


\begin{table}[hbt!]
    \centering
     \vspace*{-0.35cm}
    \caption{Object and hologram plane specification}
    \label{table:specification}
    \includegraphics[width=8cm, height=3.5cm]{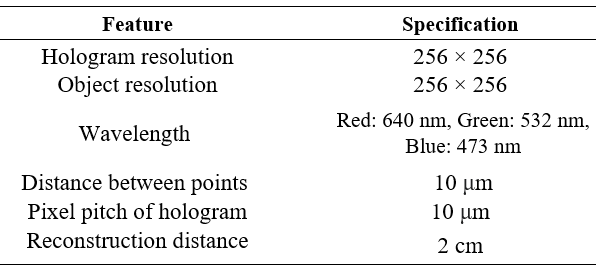}
\end{table}

Our DWT hologram generation algorithm can be summarized as follows:\\
1) In the first stage, we calculate 3 levels of wavelet for our object using Haar wavelet filters. The first level of DWT consists of $L$, $H$, $V$, and $D$. Similarly, the level of $L$ on its own is decomposed into $LL$,  $LH$, $LV$, and $LD$. Finally, the level of $LL$ is decomposed into $LLL$, $LLH$, $LLV$, and $LLD$. The point is, the size of each new level is 4 times smaller than the previous one and requires less computation for hologram formation. 

\begin{figure}[hbt!]
    \centering
    \includegraphics[scale= 0.7]{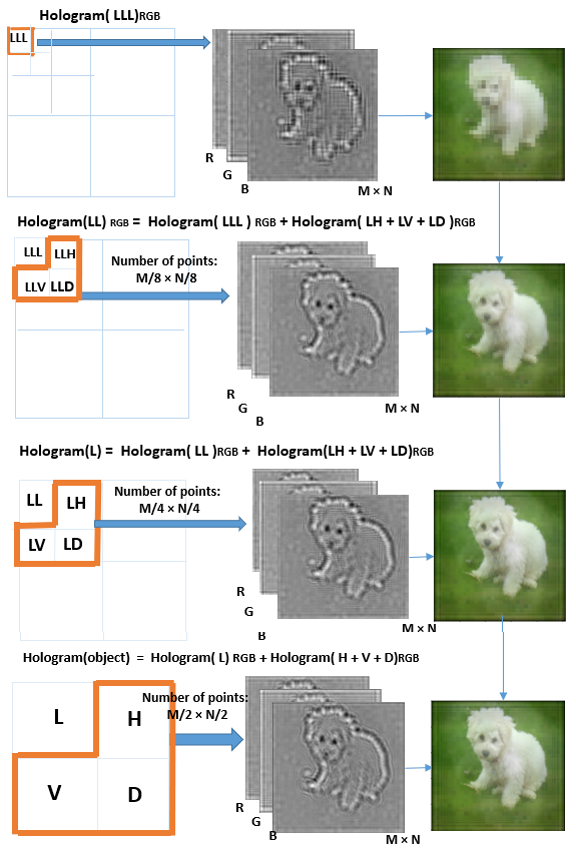}
    \caption{Block diagram of DWT outputs and calculation of the residual map with their sizes at each level.}
    \label{fig:Diagram}
\end{figure}

\begin{figure}[hbt!]
    \centering
    \includegraphics[scale = 0.8]{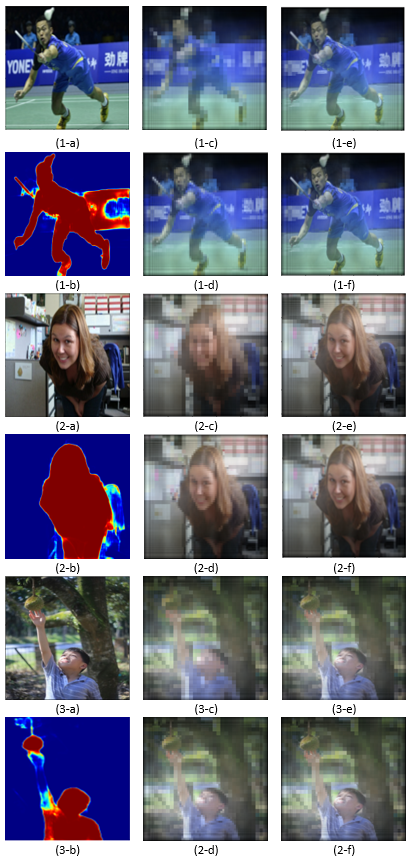}
    \caption{Three original objects with indices of 1, 2, and 3 (a) are shown. 1, 2, 3 (b) are saliency maps by PoolNet \cite{PoolNet}. 1, 2, 3 (c-e) show reconstructed objects of levels of $LLL$, $LL$, $L$, and the object level, respectively.}
    \label{fig:result}
\end{figure}

\begin{figure}
    \centering
    \includegraphics[width=8.3cm]{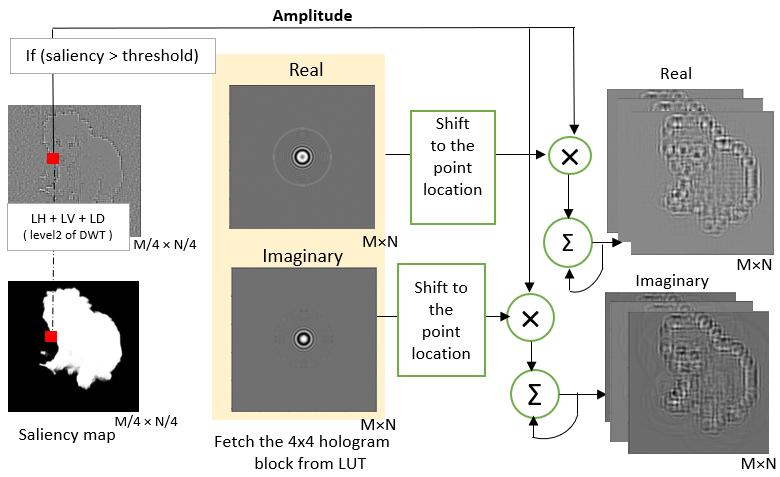}
    \caption{Block diagram of hologram computation based on saliency map at second level of DWT with $1/4$ resolution.}
    \label{fig:Loopback}
\end{figure}

We also quantize our saliency map into 2, 4, and 8 times smaller sizes by taking the maximum per block. As a matter of fact, the sizes of saliency maps correspond to wavelet outputs at each level. Fig. \ref{fig:result}(1,2,3)-b shows the saliency map before quantization. 
\\2) As shown on the first row of Fig.\ref{fig:Diagram}, for the third level of DWT i.e. $LLL$, we calculate the hologram using $8\times8$ LUT hologram block. For each object point of $LLL$, we shift $8\times8$ LUT hologram block to the point location and then multiply it by the intensity of that point. We do the same for all $LLL$  points with a saliency level of more than 0; meaning that all foreground and background blocks are included. Reconstructed holograms are shown in Fig. \ref{fig:result}(1,2,3)-c. 
Obviously, users can see a silhouette of the object but with a low resolution, similar to how progressive JPEG operates \cite{jpeg2000}.\\
3) As shown in the second row of Fig. \ref{fig:Diagram}, when we want to improve the resolution on more salient areas, we have to proceed to the second level of DWT, i.e. from $LLL$  to $LL$. 
Hence, to reach $LL$, we calculate the summation of $LLH$, $LLV$, and $LLD$ and call it the residual map which is the difference between two levels of $LLL$  and $LL$. 
Then, we calculate the hologram for this residual map. The hologram of the residual map can take our previous hologram to the hologram level of $LL$ with a higher resolution.  We calculate the hologram of this residual map similar to the previous level. Fig. \ref{fig:Loopback} shows hologram computation of all points on the residual map at the second level with $4\times4$ LUT hologram blocks, real and imaginary. We do this only for blocks with a saliency of more than a threshold, $t_1$. By adding their hologram to the previous one, it takes the hologram to the second level of resolution which is equivalent to having $LL$ at some parts and still having $LLL$ in the background. Fig.\ref{fig:result}(1,2,3)-e shows the reconstructed result.
\\
4) We do the same to reach the $L$ level as the third row of Fig. \ref{fig:Diagram} shows. Fig,\ref{fig:result}(1,2)-d also displays the corresponding reconstructed image, which is the first level of resolution. This level also is done to compensate the resolution only for points with a middle level of saliency, more than threshold $t_2$.\\
5) At the last stage, we need to have the highest hologram resolution for points with the highest saliency level by using $1\times1$ hologram LUT blocks. Hence, by calculating the addition of D, H, and V, and then calculating the hologram of this residual map, only on point with saliency more than $t_3$, we reach the highest quality for them. In Fig.\ref{fig:result}(1,2,3)-f, the front object is reconstructed purely. 

Eventually, our method can be extended to any 3D scene object provided that we replace this saliency detector with a 3D saliency detection method and obtain levels of wavelet for the 3D object and have LUT hologram blocks for different depths. Note that, in this paper, our reconstruction method is based on the inverse Angular spectrum approach \cite{AngularSpec}, using the same parameters of the scene. 

\section{EXPERIMENTAL RESULTS}
In this section, we show the number of operators and quantitative results at each level, for objects shown in Fig. \ref{fig:result}, which are samples of the ECSSD database \cite{Dataset_ECSSD}. 
To show the efficiency of our method, in table \ref{table:computations} we show the number of operators for a single color channel of objects, at each level. we define an operand including three steps of multiplication, shift, and addition as shown in Fig. \ref{fig:Loopback}. 
The three first columns show the number of operators needed for three levels of wavelet which are cumulative with their previous levels. The first column indicates that 1024 operators are used to calculate all $8 \times 8$ hologram blocks; results are shown in Fig. \ref{fig:result}(1,2,3)(c). In the following columns, although operational numbers can vary in different objects based on their saliency map, our proposed method can highly reduce hologram computations at each level, e.g. Fig. \ref{fig:result}(1,2,3)(c-e). Note that, the fourth column shows the number of operators from $L$ to the object level; results are shown in Fig. \ref{fig:result}(1,2,3)(f). The last column also shows a calculated fully point-wise hologram with LUT blocks of $1 \times 1$ for the object without decomposition, where $M \times N$ operators are needed which here is 65,536 for $256 \times 256$ resolution. This reconstructed object is the most accurate hologram for a scene parameter. Hence, we compare our results with this reconstructed result, qualitatively in Table \ref{table:ssim}. Additionally, comparison with this also makes the SSIM parameter free of scene parameters and scales. 
In this table, we show the progress of SSIM for different levels of wavelet in our proposed method, in comparison to a fully point-wise reconstructed hologram. Clearly, $LLL$  has the lowest SSIM, and the object level has the highest one. 
According to the two above tables, we can interpret that the number of operators for $L$ for the object level is relatively high whereas its SSIM is relatively close to the level $L$. This closeness can be visually realized in Fig. \ref{fig:result}. Hence, in such a case we prefer our method to end up to level $L$ for both high quality and efficiency. However, even if our proposed method proceeds to the object level, the cumulative number of operators is still less than half of the operator of the original object made for all object points (fourth column compared with the last one).
In fact, the advantage of our proposed method is that some points suffice with level $LLL$, the remaining suffice on $LL$ and others are in need of $L$ level which leads to high qualitative results on salient objects. 
In this paper, threshold parameters of $t_1$, $t_2$, $t_3$ are 0.9, 0.7, 0.5, 0 for low to high levels of wavelet on the saliency map.




\begin{table}
\centering
 \vspace*{-0.4cm}
\caption{Cumulative Number of operators at each level.}\label{table:computations}
\includegraphics[width=8.8cm]{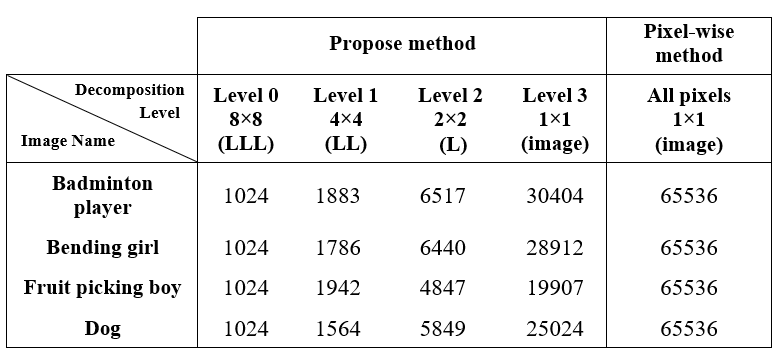}
\end{table}

\begin{table}[htb!]
\centering
\vspace*{-0.4cm}
\caption{Quality assessment based on SSIM metric.}\label{table:ssim}
\includegraphics[width=8cm]{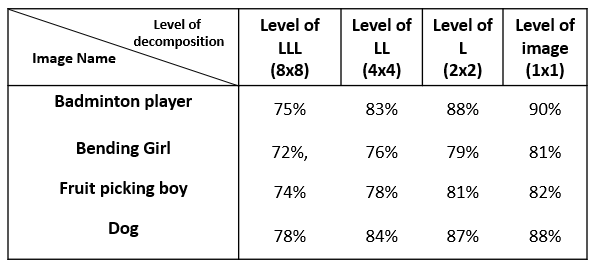}
\end{table}
\vspace*{-0.8cm}
\section{Conclusion}
\vspace*{-0.2cm}
In this paper, we speed up the hologram computation by decomposing our object points to three levels of wavelet. We made our final hologram (L) with a higher level of wavelet on less salient points and lower levels of wavelet on more salient ones. We applied our method to a number of 2D objects and show that not only the number of operators for summation of all three levels are lower than point-wise method but also SSIM stays relatively high.




\bibliographystyle{IEEEbib}
\bibliography{refs}

\end{document}